\documentclass[12pt,prd,nofootinbib]{revtex4}

\def\be{\begin{equation}}
\def\ee{\end{equation}}
\begin{document}
\title{Generating Gowdy cosmological models}
\author{Alberto S\'anchez, Alfredo Mac\'\i as}
\email{asanchez@nuclecu.unam.mx, amac@xanum.uam.mx}
\affiliation{Departamento de F\'isica,
Universidad Aut\'onoma Metropolitana--Iztapalapa\\
Apartado Postal 55--534, C.P. 09340, M\'exico, D.F., M\'exico}
\author{Hernando Quevedo}
\email{quevedo@physics.ucdavis.edu}
\affiliation{Instituto de Ciencias Nucleares\\
Universidad Nacional Aut\'onoma de M\'exico\\ 
A.P. 70-543,  M\'exico D.F. 04510, M\'exico\\}
\affiliation{Department of Physics \\
University of California \\
Davis, CA 95616}

\begin{abstract}

Using the analogy with stationary axisymmetric solutions,
we present a method to generate new analytic cosmological
solutions of Einstein's equation belonging to the class of
$T^3$ Gowdy cosmological models. We show that the solutions
can be generated from their data at the initial singularity
and present the formal general solution for 
arbitrary initial data. We exemplify the method by constructing
the Kantowski-Sachs cosmological model and a
generalization of it that corresponds to an unpolarized $T^3$ Gowdy
model. 
\end{abstract}
\maketitle
\section{Introduction}

Gowdy metrics \cite{Gowdy,Gowdy3}, which are exact
solutions of  Einstein's vacuum field
equations, represent cosmological models
with various possible topologies ($S^3\,,S^1\times S^2\,, T^3$)
and are spatially compact inhomogeneous spacetimes admitting
two commuting spacelike Killing vector fields. 
They are of especial importance for the study of formation
of singularities in general relativity. 
Many efforts \cite{varios1,vel,varios2,Berger}
have been made in order to derive and analyze this type of
solutions, especially by using numerical methods. The 
physical structure of these metrics is sufficiently 
simple to expect that the singularities can be analyzed
in detail, but their mathematical structure is sufficiently
complicated so that the global dynamical behavior is still far from
being completely understood. One of the most intriguing
questions concerns the structure of the curvature singularity 
that is expected to appear at certain spacelike boundary 
of the associated spacetime. Many studies have been devoted
to the so-called ``asymptotically velocity term dominated"
(AVTD) behavior which states that near the singularity 
each point in space is characterized by a different
spatially homogeneous cosmology \cite{eardley}.
The idea of AVTD behavior was originally proposed
in \cite{bkl} more than 30 years ago, but is still
being debated. Numerical analysis of the Gowdy models 
have shown that they do become AVTD near the singularity,
except at a set of isolated points, where there are ``spikes"
in the behavior of the metric functions. The origin of
these spikes is investigated in \cite{Berger}. 
Gowdy metrics have been also analyzed as toy models
in quantum midisuperspace gravity \cite{quantum}.

Numerical methods have been extensively used to 
investigate Gowdy models, but only 
recently it has been argued that solutions generating
techniques can be applied in this case \cite{samos,her,rendall}
to generate new solutions and that even a ``simple change of coordinates''
can be applied to reinterpret certain stationary axisymmetric solutions
as $S^1\times S^2$ Gowdy cosmological models \cite{prd}.
The reason why these methods can be used also in this
case is due to the well-known fact that the solution
generating techniques are applicable to any spacetime
which admits two (commuting) Killing vector fields. 
In this work,
we will concentrate on $T^3$ Gowdy cosmological models
and will see that a complex coordinate transformation, together
with a complex change of metric functions, allows us to apply
in a straightforward manner the well-known solution generating
techniques that have been
intensively used for stationary axisymmetric solutions.

This paper is organized as follows. In Section \ref{sec2} we
derive the ``transformation'' that relates stationary
axisymmetric solutions with Gowdy $T^3$ models.
We will show that due to this analogy, the AVTD behavior
in Gowdy $T^3$ is mathematically equivalent to the
behavior of stationary asixymmetric solutions near
the axis of symmetry. In Section \ref{sec3} we show that
Sibgatullin's method for constructing solutions
can be applied in the case of Gowdy $T^3$ models
and in Section \ref{sec4} we present several examples
of exact solutions generated by using this method.
Finally, Section \ref{con} is devoted to the conclusions and some
remarks about different possibilities of generalizing
the results derived in this work.

\section{Stationary axisymmetric solutions and Gowdy $T^3$ models}
\label{sec2}

Consider the line element for stationary axisymmetric spacetimes
in the Lewis-Papapetrou form \cite{kramer}
\begin{equation}
ds^2 = - e^{2\psi} (dT + \omega d\phi)^2 + e^{-2\psi} [e^{2\gamma}(d\rho^2 +
dz^2)
+ \rho^2 d\phi^2]  \ ,
\label{star}
\end{equation}
where $\psi,\ \omega,$ and $\gamma$ are functions of the nonignorable
coordinates
$\rho$ and $z$. The ignorable coordinates $T$ and $\phi$ are associated with
the two Killing vector fields $\eta_I = \partial/\partial T$ and
$\eta_{II} = \partial/\partial \phi$. The field equations take the form
\begin{equation}
\psi_{\rho \rho} + {1\over \rho}\psi_\rho + \psi_{zz} +
{e^{4\psi}\over 2\rho^2}(\omega_\rho^2 + \omega_z^2) = 0 \ ,
\label{psi}
\end{equation}
\begin{equation}
\omega_{\rho \rho} - {1\over \rho}\omega_\rho + \omega_{zz} +
 4 (\omega_\rho\psi_\rho + \omega_z\psi_z ) = 0 \ ,
\label{omega}
\end{equation}
\begin{equation}
\gamma_\rho = \rho(\psi_\rho^2 - \psi_z^2) -
{e^{4\psi}\over 4 \rho^2}(\omega_\rho^2 -  \omega_z^2)\ ,
\label{gammarho}
\end{equation}
\begin{equation}
\gamma_z = 2\rho \psi_\rho\psi_z - {1\over 2\rho} e^{4\psi} \omega_\rho\omega_z
\ ,
\label{gammaz}
\end{equation}
where the lower  indices represent partial derivative with respect 
to the corresponding coordinate. 

Consider now the following coordinate transformation $(\rho, t)\rightarrow
(\tau,\sigma)$
and the complex change of coordinates $(\phi,z)\rightarrow (\delta,\chi)$
defined
by
\begin{equation}
\rho= e^{-\tau}, \quad T=\sigma, \quad z=i\chi, \quad \phi = i\delta,
\label{trans1}
\end{equation}
and introduce the functions $P$, $Q$ and $\lambda$ by means of the relationships
\begin{equation}
\psi ={1\over 2}(P-\tau), \quad Q = i\omega, \quad \gamma = {1\over 2}
\left(P-{\lambda\over 2} -{\tau\over 2}\right)\ .
\label{trans2}
\end{equation}
Introducing Eqs.(\ref{trans1}) and (\ref{trans2}) into the line element
(\ref{star}),
we obtain
\begin{equation}
-ds^2= e^{-\lambda/2} e^{\tau/2} ( - e^{-2\tau} d\tau^2 + d\chi^2)
+ e^{-\tau} [e^P (d\sigma + Q d\delta)^2
+ e^{-P} d\delta^2] \ .
\label{t3}
\end{equation}
Let us take $\tau\geq 0$ (what seems reasonable in
virtue that the radial coordinate $\rho = e^{-\tau} \geq 0$) and
``compactify'' the new coordinates as  $0 \leq \chi,\ \sigma, \ \delta
\leq 2\pi$ (a less
reasonable condition since in general $ -\infty < z < +\infty$ and
$T\geq 0$).
The line element (\ref{t3}) with the coordinates $\tau,\ \chi, \sigma$
and $\delta$ in the range given above is known
 as the line element
for Gowdy $T^3$ cosmological models \cite{Berger}.
Furthermore, one can verify by direct calculation
that the action of the
transformations (\ref{trans1}) and
(\ref{trans2}) on the field equations (\ref{psi})-(\ref{gammaz}) yields exactly
the field equations for the Gowdy cosmological models which after some
algebraic manipulations can be written as
a set of two second order differential equations for $P$ and
$Q$
\begin{equation}
P_{\tau\tau} - e^{-2\tau} P_{\chi\chi} - e^{2P}(Q_\tau^2 -
e^{-2\tau}Q_\chi^2) = 0 \ ,
\label{t3eqp}
\end{equation}
\begin{equation}
Q_{\tau\tau} - e^{-2\tau} Q_{\chi\chi} + 2(P_\tau Q_\tau -
e^{-2\tau}P_\chi Q_\chi) = 0 \ ,
\label{t3eqq}
\end{equation}
and two first order differential equations for $\lambda$
\begin{equation}
\lambda_\tau = P_\tau^2 + e^{-2\tau}P_\chi^2 + e^{2P} (Q_\tau^2 +
e^{-2\tau}Q_\chi^2) \ ,
\label{t3eqlam1}
\end{equation}
\begin{equation}
\lambda_\chi = 2(P_\chi P_\tau +
e^{2P}Q_\chi Q_\tau)  \ .
\label{t3eqlam2}
\end{equation}
It should be emphasized that
this method for ``deriving'' the Gowdy line element from the stationary
axisymmetric one
involves real as well as complex transformations at the level
of coordinates and
metric functions. It is, therefore, necessary to demand that the
resulting metric
functions $P, \ Q,$ and $\lambda$ be real. That means that in general it is
not possible to take an axisymmetric stationary solution and apply the
transformations to obtain a Gowdy cosmological model. If the resulting
functions are not real, they cannot be physical reasonable
solutions to the real equations (\ref{t3eqp})--(\ref{t3eqlam2}).
These transformations can be used only as a guide to get some insight into
the form of the new solutions. In any case, the corresponding
field equations have to be invoked in order to confirm the correctness
of the solution.

A very useful form for analyzing the field equations of spacetimes
with two (commuting) Killing vector fields is the Ernst representation.
In fact, this convenient form for the field equations was first proposed
for axisymmetric stationary solutions \cite{ernst}, but since
then it has been applied in many different configurations 
\cite{grif,nora1,nora2}.
Here we will present the Ernst representation of the main field
equations (\ref{t3eqp}) and (\ref{t3eqq}) which is especially adapted
to the coordinates used here.
To this end, let us introduce a new coordinate  $t$ and a new function
$R=R(t,\chi)$
by means of the equations \cite{samos,prd}
\be
t=e^{-\tau}, \quad
R_t = t e^{2P} Q_\chi  ,\quad  R_\chi = t e^{2P} Q_t .
\label{defr}
\ee

Then, the field equation (\ref{t3eqp}) can be expressed as
\begin{equation}
t^2\left( P_{tt} + {1 \over t} P_t - P_{\chi\chi} \right)
+ e^{-2P} (R_t^2 - R_\chi^2) = 0 \ ,
\label{eqpr}
\end{equation}
whereas Eq.(\ref{t3eqq}) for the function $Q$ turns out to be
equivalent to the the integrability condition $R_{t\chi}=R_{\chi t}$. 
However, an alternative and convenient
equation is obtained by introducing Eq.(\ref{defr}) 
directly into Eq.(\ref{t3eqq}). So we obtain
\begin{equation}
t e^P \left( R_{tt} + {1 \over t} R_t - R_{\chi\chi} \right)
- 2 [(t e^P)_t R_t - (te^P)_\chi R_\chi ] = 0  \ ,
\label{eqqr}
\end{equation}
an equation which of course becomes an identity if the
integrability condition $R_{t\chi} = R_{\chi t}$ is satisfied.
We can now introduce the complex Ernst potential $E$ and
the complex gradient operator $D$ as
\begin{equation}
E = t e^P + i R \ , \qquad {\rm and} \qquad
D= \left({\partial \over \partial t} \ , \
         i {\partial \over \partial \chi} \right) \ ,
\label{ernstpot}
\end{equation}
which allow us to write the main field equations in the
{\it Ernst-like representation}
\begin{equation}
Re(E )\left(D^2 E + {1\over t} D t\, D E \right)
 - (D  E )^2 = 0 \ .
\label{ernstt3}
\end{equation}
It is easy to verify that the field equations (\ref{eqpr}) and
(\ref{eqqr})
can be obtained as the real and imaginary part of the Ernst
equation (\ref{ernstt3}), respectively. For the sake of
completeness, we rewrite the system of first order, partial,
differential equations (\ref{t3eqlam1}) and (\ref{t3eqlam2})
in terms of the Ernst potential: 
\be
\lambda_t = -{t\over 2} \left( C_+ C_+^*  + C_- C_-^*\right) \ ,
\label{lamt}
\ee
\be
\lambda_\chi = -{t\over 2} \left( C_+ C_+^*  - C_- C_-^*\right) \ ,
\label{lamtheta}
\ee
where
\be
C_\pm = {1\over {\rm Re} (E)} \left(E_t \pm E_\chi\right) 
- {1\over t} \ ,
\ee
and the asterisk denotes complex conjugation.

If the Ernst potential $E$ is known, then it is easy to recover
the metric functions $P,\ Q$ and $\lambda$
which enter the line element (\ref{t3})
of Gowdy $T^3$ cosmological models. In fact, from Eq.(\ref{ernstpot})
one can algebraically construct the functions $P$ and $R$. Then
the function $Q$ can be obtained by solving the system of two first
order partial differential equations given in (\ref{defr}). Notice
that the integrability condition of this last system is satisfied
by virtue of Eq.(\ref{ernstt3}). 
Finally, the system (\ref{lamt}) and (\ref{lamtheta}) for the
function $\lambda$ can
be solved by quadratures since its integrability condition coincides with
the Ernst equation (\ref{ernstt3}). Consequently, all the information
about any Gowdy $T^3$ cosmological model is contained in the
corresponding Ernst potential.

One of the most important properties of the Ernst representation
(\ref{ernstt3}) is  that it is very appropriate to investigate
the symmetries of the
field equations. In particular, the symmetries of the Ernst equation
for stationary axisymmetric spacetimes have been used to develop the
modern solution generating techniques \cite{dh}, like the B\"acklund method,
Belinsky-Zakharov inverse scattering method, the
Hoenselaers-Kinnersley-Xanthopoulos method, and others (for
an introductory review and detailed references see \cite{fort}).
In all these methods it is necessary to start from a given
``seed'' solution which has to be specified in the whole spacetime
(except, perhaps, in the regions where the metric possesses true
curvature singularities). An alternative approach for exploring
the symmetries inherent in the Ernst equation was explicitly
developed by Sibgatullin \cite{sib} and consists in constructing
exact solutions to the Ernst equation from initial data specified
only on certain hypersurface (submanifold) of the spacetime.
For instance, in the case of stationary axisymmetric spacetimes,
Sibgatullin's method allows one to construct exact solutions
from their data on the axis of symmetry. In the following sections
we will show that Sibgatullin's method can be applied in the
case of Gowdy cosmological models and will present several examples
of its application.

\section{Constructing solutions from AVTD data}
\label{sec3}

As we have mentioned above, an important property of Gowdy
cosmological models is its AVTD behavior near the initial
singularity. In the case of $T^3$ models it can be shown
that the singularity is approached in the limit
$\tau \rightarrow \infty$. The AVTD behavior implies
that at the singularity all spatial derivatives of the
field equations can be neglected and only the temporal
behavior is relevant. On the other hand,
the transformation (\ref{trans1}) indicates that the limit
$\tau \rightarrow \infty$ is equivalent
to the limit $\rho \rightarrow 0$; however,
this is true only at the level of coordinates and
a more detailed analysis is necessary to make
sure that this analogy is also valid at the
level of explicit solutions. To this end,
let us consider the system of partial differential
equations for $\psi$ and $\omega$ given in
Eqs.(\ref{psi}) and (\ref{omega}). If we neglect
the spatial dependence on $z$, which accoring
to the transformation (\ref{trans1}) is equivalent
to the spatial dependence on $\chi$ in Gowdy models,
then we obtain
a system of differential equations which can be
solved by quadratures and yields
\be
\psi = {1\over 2} \ln [ a(\rho^{1+c} + b^2 \rho^{1-c}) ]
\ ,\qquad
\omega= {i b\over  a(\rho^{1+c} + b^2 \rho^{1-c}) } + i d \ ,
\label{assol}
\ee
where $a,\ b, \ c$ and $d$ are arbitrary real functions of $z$.
Clearly, this solution is meaningless when considered as
a stationary axisymmetric spacetime. However, if we now follow
the prescription given in Eqs.(\ref{trans1}) and (\ref{trans2})
for obtaining Gowdy models, we fill find that solution
(\ref{assol}) ``corresponds'' to the Gowdy model
\be
P = \ln [ a( e^{-c\tau} + b^2 e^{c\tau})]\ , \qquad
Q = {b \over a(e^{-2c\tau} + b^2)} + d \ ,
\label{avtd}
\ee
where now $a ,\ b, \ c$ and $d$ are to be considered
as arbitrary real functions of the coordinate $\chi$.
It is straightforward to verify that the expressions given
in Eq.(\ref{avtd}) satisfy the Gowdy field equations
(\ref{t3eqp}) and (\ref{t3eqq}) in its ``truncated'' form,
i.e., when the spatial derivatives are neglected.
The solution (\ref{avtd}) is known in the literature
as the AVTD solution for Gowdy $T^3$ models \cite{Berger}
and dictates the behavior of these models near the
singularity $\tau\rightarrow \infty$. Thus, we have
``derived'' the AVTD solution starting from its
stationary axisymmetric counterpart. This is a
further indication that the behavior of Gowdy models
at the initial singularity is mathematically equivalent to the
behavior of stationary axisymmetric solutions at the axis.
For the sake of completeness we also quote here the
value of the function $\lambda$ corresponding to the
AVTD solution (\ref{avtd}) that can be obtained by
integrating Eq.(\ref{t3eqlam1}): 
\be
\lambda = \lambda_0 - c^2 \ln t \ , 
\label{lambdaavtd}
\ee
where $\lambda_0$ is an additive constant. 
Furthermore, the corresponding AVTD Ernst potential
can be obtained by introducing Eq.(\ref{avtd}) into 
from Eqs.(\ref{ernstpot}) and (\ref{ernstt3}). Then 
\be
E = a[e^{-(1+c)\tau} + b^2 e^{-(1-c)\tau}] + i R^{avtd} \qquad {\rm with}
\qquad R^{avtd}_\chi = - 2 a b c \ .
\label{ernstavtd}
\ee
If we define 
\be
E(\tau\rightarrow\infty,\chi) = e(\chi) \ 
\label{ernstsing}
\ee
as the Ernst potential at the singularity, we see from Eq.(\ref{ernstavtd}) 
that for $c\in (-1,1)$ only the imaginary part remains, 
$e(\chi)=iR^{avtd}$. This means 
that the real part of $e(\chi)$ is arbitrary and since $R^{avtd}$ is given
in terms of the real part it is also arbitrary. If $c \notin (-1,1)$,
the Ernst potential diverges at the singularity for arbitrary values 
of the functions $a$ and $b$. In the limiting case $c=\pm 1$, the 
Ernst potential at the singularity is regular, but again no 
conditions appear for the behavior of the functions $a$ and $b$.
Consequently, the AVTD behavior does not impose any conditions 
on the function $e(\chi)$. We will now see that it is possible 
to use this function to construct the corresponding Ernst potential
$E(\tau,\chi)$.

Sibgatullin's method  \cite{sib} has been developed to construct
exact stationary axisymmetric solutions starting from their
data on the axis of symmetry. It is based upon the fact that 
the Ernst equation possesses symmetry properties associated with
an infinite-dimensional Lie group which transforms one solution
of this equation into another solution of the same equation. 
This implies remarkable analyticity properties that make it
possible to reduce the Ernst equation to a system of linear
integral equations which can be integrated explicitly if
initial data is known, for instance, on the axis of symmetry. 
It is clear that the Ernst-like representation (\ref{ernstt3}) 
possesses similar symmetry properties. On the other hand,
we have shown that the behavior of stationary axisymmetric
solutions near the axis is mathematically equivalent to 
the behavior of Gowdy $T^3$ cosmological models near the
singularity.
Thus, it should be possible to  construct Gowdy
cosmological models starting from the value of the corresponding
Ernst potential at the singularity. It turns out that 
Sibgatullin's method can be generalized in a straightforward
manner to include the case of Gowdy models. A detailed explanation
of the procedure necessary to obtain the system of linear
integral equations associated with the Ernst equation 
is  given in \cite{sib}. Here we will only quote
the main steps of the construction. Assume that the value of
the Ernst potential is known at the initial singularity, i.e.
$e(\chi)$ is given.
 Then, the Ernst potential can be generated by means
of the integral equation
\begin{equation}
\label{ecu49}
E(t,\chi)=\frac{1}{\pi} \int_{-1}^{1}
\frac{ e(\xi) \mu(\xi)  }
{ \sqrt{1-s^2} } \, ds \,,
\end{equation}
where the unknown function $\mu(\xi)$ has to be found from
the singular integral equation
\be
\label{ecu50}
\int_{-1}^{1} \frac{\mu({\xi})[e^*({\eta})+e({\xi})]}
{({s} -{\kappa})\sqrt{1-{s^2}}}\, ds =0\,,
\ee
with the normalization condition
\be
\label{ecu51}
\int_{-1}^{1}
\frac{{\mu}({\xi})}{\sqrt{1-{s^2}}}\, ds =\pi\,,
\ee
where $\xi= \chi + t s$, $\eta = \chi + t \kappa$, with
$s,\kappa \in [-1,1]$. 

Notice that for this method no condition is imposed on
the behavior of $e(\chi)$. This is in accordance with 
the result obtained above about the AVTD behavior of 
the Ernst potential near the singularity. Once $e(\chi)$
is given in any desired form, one only has to calculate
the integral (\ref{ecu49}) to find the Ernst potential.
However, to calculate this integral one first has to 
find the function $\mu(\xi)$ by means of the singular 
equation (\ref{ecu50}) and the normalization condition 
(\ref{ecu51}). In practice, for a given $e(\xi)$
one has to make a reasonable ansatz for $\mu(\xi)$ 
such that it allows the existence of solutions
for the integral singular equation (\ref{ecu50}).

\section{Examples of Gowdy $T^3$ models}
\label{sec4}

The cases where the Ernst potential at the initial singularity
behaves as a rational function are relatively easy to analyze. In
this section we will present two such examples. Let us consider
the following simple example of an Ernst potential at the
singularity 
\be e(\chi) = {\chi_0 - \chi\over \chi_0 + \chi} \ ,
\label{ex1} 
\ee 
where $\chi_0$ is a real constant. 
The first step of the construction is to find the unknown function
$\mu$ according to Eqs.(\ref{ecu50}) and (\ref{ecu51}). A reasonable
ansatz is again a rational function \cite{sib}
\be
\mu= A_0
+ {A_1\over \xi-\xi_1} \ , \label{muks}
\ee
where $\xi_1$  is the root of the equation 
$e(\xi)+\tilde e (\xi) = 0$ (in this case $\xi_1=\chi_0$)
  and $A_0,\ A_1,$ are
functions of $t$ and $\chi$. To handle the integrals which follow
from the singular integral equation we use the following standard
formulae
\begin{eqnarray}
\label{ecu67}
 \int_{-1}^{1}
\frac{d s }{\sqrt{1-s^2}}&=&\pi \ ,\\
\label{ecu68}
\int_{-1}^{1} \frac{ d s }{(a+isb) \sqrt{1-s^2}}&=&
\frac{\pi}{\sqrt{a^2+b^2}}\ ,\\
\label{ecu69} \int_{-1}^{1} \frac{ d s }{(s-\kappa)(s-\gamma)
\sqrt{1-s^2}}&=& \frac{\pi}{(\kappa-\gamma)\sqrt{\gamma^2-1}} \ ,
\end{eqnarray}
where $a$, $b$ and $\gamma$ are arbitrary constants.

Introducing Eq.(\ref{muks}) into the normalization condition (\ref{ecu51}),
we obtain
\be
A_0 + {A_1\over r_-} = 1 \ ,
\label{normks}
\ee
where $r_- = \sqrt{(\chi-\chi_0)^2 - t^2}$, whereas the integral singular
equation (\ref{ecu50}) leads to 
\be
A_0-\frac{A_1 ( r_{+}+r_{-})} {2\chi_0 r_{-}} =0  \ .
\label{seqks}
\ee
The last two equations can be used to find the explicit values of 
$A_0$ and $A_1$ which then can be replaced in the result of the 
integration of Eq.(\ref{ecu49}) and yield 
\be
E(t,\chi) = - A_0  - {A_1-2\chi_0 A_0\over r_+}
=  {2\chi_0 - r_+ - r_- \over 2\chi_0 + r_+ + r_- } \ ,
\label{ernstks}
\ee
where $r_+ = \sqrt{(\chi+\chi_0)^2-t^2}$.
It is easy to check that indeed this is a solution to the
Ernst equation (\ref{ernstt3}). Since the resulting Ernst
potential is real, the solution corresponds to a
polarized ($Q=0)$ Gowdy model. The expression for the
metric function $P$ can easily be obtained from the
definition (\ref{ernstpot}) and Eq.(\ref{ernstks}),
and the remaining function $\lambda$ can be calculated
(up to an additive constant) by quadratures
from Eq.(\ref{lamt}) and (\ref{lamtheta}):
\be
\lambda= \ln\left[{1\over t}{(r_+ r_-)^2\over
                   (r_++r_-+ 2\chi_0)^4}\right] \ .
\label{lamks}
\ee
The physical significance of this solution becomes
plausible in a different system of coordinates which
we introduce in two steps.
Let us first introduce in the ($\tau, \chi$)-sector
of the line element (\ref{t3})
coordinates $x$ and $y$
by means of the relationships
\be
e^{-2\tau}= t^2 = \chi_0^2 (1-x^2)(1-y^2)\ , \qquad
\chi = \chi_0x y \ ,
\ee
or the inverse transformation law
\be
x = {r_++r_-\over 2\chi_0}\ , \qquad
y = {r_+-r_-\over 2\chi_0}\ ,
\label{xy}
\ee
so that the metric functions become
\be
P = \ln\left[{1-x\over 
\chi_0\sqrt{(1-x^2)(1-y^2)} (1+x)}\right]\ , \quad 
\lambda= \ln\left[{ (x^2-y^2)^2\over 
\chi_0 \sqrt{(1-x^2)(1-y^2)}(1+x)^4}\right]\ .
\ee
The second transformation affects now all the
sectors of the line element (\ref{t3}) and is
defined by
\be
x= {T\over \chi_0} - 1\ , \quad
y=\cos\theta \ ,\quad \sigma= r\ ,\quad \delta=\phi\ .
\ee
Then, after some algebraic manipulations, the metric
can be written as
\be
-ds^2 = -\left({2\chi_0\over T} - 1\right)^{-1} dT^2 
+ \left({2\chi_0\over T} - 1\right) dr^2 +
T^2 (d\theta^2+\sin^2\theta d\phi^2) \ ,
\label{ks}
\ee
an expression that can immediately be recognized
as the Kantowski-Sachs cosmological model \cite{ks,kras}.
Thus, we have shown that the Kantowski-Sachs metric
can be constructed from the value of its Ernst 
potential at the singularity (\ref{ex1}). 

Consider now the more general case
\be 
e(\chi) = {\chi_0 - \chi -i\chi_1 \over \chi_0 + \chi + i\chi_1} \ ,
\label{ex2}
\ee
where $\chi_0$ and $\chi_1$ are real constants. 
The unknown function $\mu(\xi)$ can be sought in the 
form 
\be
\mu = A_0 + {A_1\over \xi - \xi_1 } + {A_2\over \xi - \xi_2 } \ ,
\label{mukg}
\ee
where $\xi_{1,2} = \pm \alpha = \pm \sqrt{\chi_0^2 - \chi_1^2}$
are the roots of the equation $e(\xi)+e^* (\xi) = 0$.
Substituting Eq.(\ref{mukg}) in the integral equation (\ref{ecu50})
we obtain the system 
\begin{eqnarray}
-A_0+\frac{A_1}{\chi_0+i\chi_1+\alpha}+
\frac{A_2}{\chi_0+i\chi_1-\alpha}&=&0\,,\label{as1kg} \\
-\frac{A_1 (\alpha+i\chi_1)}{r_{-} ( \chi_0+i\chi_1+\alpha)}+
\frac{A_2 (\alpha-i\chi_1)}{r_{+} ( \chi_0+i\chi_1-\alpha)}&=&0\,, 
\label{as2kg}
\end{eqnarray}
where $r_\pm = \sqrt{(\chi\pm\alpha)^2-t^2}$.
On the other hand,
the normalization condition (\ref{ecu51}) yields 
\be
A_0 + {A_1\over r_-} + {A_2\over r_+} = 1 \ ,
\ee
an equation which together with Eqs.(\ref{as1kg}) and (\ref{as2kg}) 
form a closed algebraic system that determines the coefficients of
the function $\mu$: 
\begin{eqnarray}
A_0&=&\frac{\alpha (r_{+}+r_{-})+i\chi_1(r_{+}-r_{-}) }
{\alpha (r_{+}+r_{-})+i\chi_1(r_{+}-r_{-})+2\alpha \chi_0}
\,,\nonumber \\
A_1&=&\frac{r_{-}(\alpha-i\chi_1)(\chi_0+i\chi_1+\alpha)}
{\alpha (r_{+}+r_{-})+i\chi_1(r_{+}-r_{-})+2\alpha \chi_0}\,,
\label{askg} \\
A_2&=&\frac{r_{+}(\alpha+i\chi_1)(\chi_0+i\chi_1-\alpha)}
{\alpha (r_{+}+r_{-})+i\chi_1(r_{+}-r_{-})+2\alpha \chi_0}
\,.\nonumber
\end{eqnarray}
Finally, we calculate the Ernst potential according to Eq.(\ref{ecu49})
and obtain
\begin{eqnarray}
E(t,\chi) & = & -A_0+\frac{A_1(\chi_0-i\chi_1-\alpha)}
{r_{-}(\chi_0+i\chi_1+\alpha)}+
\frac{A_2(\chi_0-i\chi_1+\alpha)}
{r_{+}(\chi_0+i\chi_1-\alpha)} \nonumber \\
  &=& {2\alpha \chi_0 - \alpha (r_++r_-) - i\chi_1(r_+-r_-)
\over 2\alpha \chi_0 + \alpha(r_++r_-) + i\chi_1(r_+-r_-)} \ .
\label{ernstkt3}
\end{eqnarray}
The calculation of the corresponding metric functions
can be carried out as described in the last section.
When integrating the systems of first order differential
equations (\ref{defr}) for $Q$ and (\ref{lamt}) and 
(\ref{lamtheta}) for $\lambda$, constants of integration
appear which we choose such that a simpler representation
is obtained in terms of the coordinates used. 
To write down the final form
of the metric functions it is convenient to 
use the coordinates ($x,\ y$) as defined in Eq.(\ref{xy})
with $\chi_0$ replaced by $\alpha$. Then 
\be
P=\ln { \chi_0^2-\alpha^2x^2- \chi_1^2y^2\over
 \alpha\sqrt{(1-x^2)(1-y^2)}[(\chi_0+\alpha x)^2+\chi_1^2y^2]}
 \ ,
\label{pkt3}
\ee
\be
Q= {2\chi_1\chi_0 (1-y^2)(\chi_0+\alpha x) \over
\chi_0^2-\alpha^2x^2- \chi_1^2y^2 } \ ,
\label{qkt3}
\ee
\be
\lambda = \ln {\chi_0^3 (x^2-y^2)^2 \over
 \sqrt{(1-x^2)(1-y^2)}[(\chi_0+\alpha x)^2+\chi_1^2y^2]^2}\ .
\label{lkt3}
\ee
This metric corresponds to an unpolarized generalization of
the Kantowski-Sachs cosmological model which is obtained
in the limiting case $\chi_1=0$. 

It is easy to see that
near the singularity ($\tau\to \infty)$ the general Ernst potential
(\ref{ernstkt3}) approaches the corresponding AVTD potential
as given in Eq.(\ref{ex2}). This could be interpreted as an 
indirect proof of the AVTD behavior of the solution obtained
here. A more direct proof can be given by analyzing the hyperbolic
velocity $v = \sqrt{P_\tau^2 + e^{2P}Q_\tau^2}$ 
(see Ref.\cite{vel}). In terms of the coordinates $x$ and $y$,
the hyperbolic velocity of the solution (\ref{pkt3}) 
and (\ref{qkt3}) is given as a rather cumbersome expression. 
Nevertheless, it is possible to perform an analysis of 
its behavior by considering the different domains of the
coordinates according to the definition equation (\ref{xy}).
One can show that in general $0\leq v < 1$ what according
to Ref.\cite{vel} implies that the solution is AVTD.

An important property of the method presented here is
that it allows to calculate an arbitrary Gowdy $T^3$ 
cosmological solution with any degree of accuracy. 
To this end, let us expand in Eq.(\ref{ecu49})
the value of the Ernst potential $e(\xi)$ in
Fourier series,
\be
e(\xi) = \sum_{k=0}^\infty e_k(t,\chi)\cos(k\varphi) \ ,
\ee
where $e_0 = e(\chi)$ and we have represented the parameter
$s$ as $s=\cos\varphi$. Then 
the unknown function $\mu(\xi)$ can be expanded in
a similar way
\be
\mu(\xi) = \sum_{k=0}^\infty \mu_k(t,\chi)\cos(k\varphi) \ .
\ee
The normalization condition (\ref{ecu51}) can easily be
calculated and implies that $\mu_0 = 1$. Furthermore,
the
general solution of the integral equation (\ref{ecu49})
can be written as \cite{sib}
\be
E(t,\chi) = e(\chi) + {1\over 2}\sum_{k=1}^\infty e_k \mu_k \ .
\ee
According to Eq.(\ref{ecu50}), the coefficients $\mu_k$ have to
satisfy the following system of algebraic equations
\be
\sum_{l=1}^\infty \mu_k (e_{k+l} - e^*_{k+l}
                       + e_{|k-l|} - e^*_{|k-l|}) = -2 e_k \ .
\ee
Thus, once the value of the Ernst potential is given at the
initial singularity [$e(\chi) = 
E(\chi, \tau\rightarrow \infty)$)]
the general solution of the Ernst equation reduces to an
infinite series with coefficients satisfying a set of pure algebraic
equations.

\section{conclusions}
\label{con}

We have shown that it is possible to generate Gowdy $T^3$ 
cosmological models starting from their data near the initial 
singularity. To this end, we first show that the Gowdy 
$T^3$ line element can be obtained from the line element
of stationary axisymmetric solutions by means of 
complex transformation that involves the metric functions
and the coordinates. The behavior of stationary axisymmetric
solutions at the axis of symmetry 
is shown to be mathematically equivalent to
the behavior of Gowdy $T^3$ models near the singularity.
In particular, we have derived the AVTD solution 
from its stationary axisymmetric counterpart. 
We then use 
the Ernst representation of the field
equations and apply Sibgatullin's method to the Ernst 
potential which can be given at the singularity 
as any arbitrary function
of the angle coordinate $\chi$.  
In particular, we have shown that the Kantowski-Sachs
cosmological model can be derived in this manner 
by starting from a specific form of the 
Ernst potential in terms of a rational function. 
We then have
found an unpolarized generalization of the 
Kantowski-Sachs cosmological model. 
This generalization has
been obtained in the same way as the Kerr metric is
obtained from its value at the axis of symmetry by using
Sibgatullin's method \cite{sib}. It is possible to consider
more general examples of Ernst potentials at the axis in terms
of rational functions. It turns out that the system of
integral equations (\ref{ecu50}) and (\ref{ecu51}) forms
a closed algebraic system from which the value of the 
function $\mu(\xi)$ can be found and the expression for
the Ernst potential can be calculated \cite{manko}. 
This method could also be applied  in the case of
Gowdy cosmological models considered here.

By expanding the value of the Ernst potential at the
singularity in terms of a Fourier series, it is 
possible to write explicitly  the general solution 
for this type of models (including the unpolarized case)
by using only a recurrence algebraic formula.
This is a result that could find some application in 
numerical investigations since it allows us to ``control''
the accuracy of the analysis by truncating the series
at any desired level.

\begin{acknowledgments}

This work was supported by CONACyT grants 42191--F and 36581--E,
DGAPA-UNAM grant IN112401, and US DOE grant DE-FG03-91ER40674.
A.S. acknowledges support from a
CONACyT fellowship. H. Q. thanks UC MEXUS for support. 
\end{acknowledgments}

\end{document}